\pdfoutput=1

\documentclass[11pt, letter]{article}
\usepackage[authoryear,round]{natbib} 
\usepackage{amsmath}
\usepackage{amssymb}
\usepackage{amsthm}
\usepackage{graphicx}
\usepackage{epstopdf}
\usepackage{epsfig}
\usepackage{url}
\usepackage{subfigmat}
\usepackage{setspace}
\usepackage{geometry}
\makeatletter
\def\@citess#1{\textsuperscript{#1)}}

\begin{document}
\title{Cryptocurrency portfolio optimization with multivariate normal tempered stable processes and Foster-Hart risk\thanks{We are grateful to Professor Svetlozar Rachev, in Texas Tech University, for his continuous guidance and encouragements on the research topic of this paper. For the first author, the opinions, findings, conclusions or recommendations expressed in this paper are our own and do not necessarily reflect the views of the Bank of Japan.}}


\author{Tetsuo Kurosaki\thanks{Bank of Japan, 2-1-1 Nihonbashi-Hongokucho, Chuo-ku, Tokyo 103-8660, Japan; \protect \\ E-mail address: tetsuokurosaki@gmail.com}, \hspace{4mm} Young Shin Kim\thanks{College of Business, Stony Brook University, Stony Brook, NY 11794-3775, United States; \protect \\ E-mail address: aaron.kim@stonybrook.edu}  
}

\date{\today}

\maketitle

\begin{abstract}
We study portfolio optimization of four major cryptocurrencies. Our time series model is a generalized autoregressive conditional heteroscedasticity (GARCH) model with multivariate normal tempered stable (MNTS) distributed residuals used to capture the non-Gaussian cryptocurrency return dynamics. Based on the time series model, we optimize the portfolio in terms of Foster-Hart risk. Those sophisticated techniques are not yet documented in the context of cryptocurrency. Statistical tests suggest that the MNTS distributed GARCH model fits better with cryptocurrency returns than the competing GARCH-type models. We find that Foster-Hart optimization yields a more profitable portfolio with better risk-return balance than the prevailing approach.
\end{abstract}






\pagebreak

\newpage
\section{Introduction}

Cryptocurrency is an entirely new finanical asset, which increases market capitalization rapidly and attracts growing attention from market participants. The advantages of cryptocurrencies over traditional currencies are abundant, and include reliability and anonymity in transactions with lower costs that are facilitated by novel blockchain technology. Despite the appeal of these advantages, cryptocurrencies present potential vulnerabilities in that they do not enjoy the aegis of central banks or any other monetary authorities. This leads to price volatility in the face of real economic events, such as the recent Covid-19 pandemic, ultimately strengthening global calls for regulations on cryptocurrency trading.\footnote{\citet{BCBS:2019} emphasizses that cryptocurrency is not legal tender and warns about the potential financial stability concerns caused by its continuous growth.} These peculiar price movements motivate scholars to study the impact of cryptocurrencies on financial markets.

From an econometric point of view, it is of interest to explore which kind of time series model accounts for highly volatile returns on and accurately forecasts risks associated with cryptocurrencies. Among expanding literature, the studies by \citet{Caporale:2019}, \citet{Cerqueti:2020}, and \citet{Troster:2019} are based on a generalized autoregressive conditional heteroscedasticity (GARCH) model. They share the common conclusion that the normally distributed GARCH model is inadequate for describing cryptocurrency returns and the introduction of non-Gaussian distribution substantially improves the goodness-of-fit of a GARCH-type model. Another approach includes the stochastic volatility model \citep{Chaim:2018}, and the generalized autoregressive score model \citep{Troster:2019}. Also, the high volatility of cryptocurrency highlights its speculative nature. \citet{Brauneis:2019} investigate the risk-return relationship of an optimized cryptocurrency portfolio based on the Markowitz mean-variance framework.

This paper studies the portfolio optimization of cryptocurrencies by employing a union of sophisticated time series models and risk measures. Four major cryptocurrencies are selected as samples. Our time series model is the multivariate normal tempered stable (MNTS) distributed GARCH model. The MNTS distribution \citep{Kim:2012} has demonstrated excellent fit to joint dynamics of physical asset returns in a number of empirical studies (\citeauthor{Anand:2016}, \citeyear{Anand:2016}; \citeauthor{Anand:2017}, \citeyear{Anand:2017}; \citeauthor{Bianchi:2019}, \citeyear{Bianchi:2019}; \citeauthor{Kim:2015}, \citeyear{Kim:2015}; \citeauthor{Kurosaki:2013a}, \citeyear{Kurosaki:2013a}; \citeauthor{Kurosaki:2013b}, \citeyear{Kurosaki:2013b}; \citeauthor{Kurosaki:2019}, \citeyear{Kurosaki:2019}; and \citeauthor{Shao:2015}, \citeyear{Shao:2015}). Our portfolio optimization strategy is based on Foster-Hart risk, which is very sensitive to risky left tail events. Foster-Hart (FH, hereafter) risk was originally introduced in the field of game theory \citep{Foster:2009}, and was subsequently applied to risk management related to financial markets (\citeauthor{Anand:2016}, \citeyear{Anand:2016}; \citeauthor{Kurosaki:2019}, \citeyear{Kurosaki:2019}; \citeauthor{Leiss:2018}, \citeyear{Leiss:2018}). These cutting-edge techniques are not yet documented in the context of cryptocurrency. Statistical tests demonstrate that the MNTS distributed GARCH model has better explanatory power for cryptocurrency returns than the normally distributed GARCH model. Also, we find that the model creates more profitable portfolios in tandem with FH risk than the traditional mean-variance approach.

The rest of this paper is organized as follows. Section \ref{Methodology} briefly introduces our methodology. Section \ref{Data and Estimation} describes the data of cryptocurrency used herein. Section \ref{Statistical Tests} outlines the empirical results of statistical tests for each cryptocurrency return. Section \ref{Portfolio Optimization} conducts portfolio optimization and discusses performance. Section \ref{Concluding Remarks} summarizes our findings.

\section{Methodology}
\label{Methodology}

We introduce the non-Gaussian time series model with the MNTS distribution as well as FH risk, which work together to achieve efficient portfolio optimization \citep{Anand:2016}. See also \ref{Supplement to Methodology} for supplementary explanations.

\subsection{Non-Gaussian time series model}

We utilize a GARCH-type model to describe the dynamics of cryptocurrency returns. Given that both autoregressive (AR) and moving average (MA) processes are typically observed in financial data, we apply the most standard ARMA(1,1)-GARCH(1,1) specification. After GARCH filtering, we obtain independent and identically distributed (i.i.d.) standard residuals $\eta_t$ with a mean of zero and unit variance for each cryptocurrency. In order to describe complicated interdependency among cryptocurrencies, we conduct multivariate modeling on each cryptocurrency's $\eta_t$ jointly. We employ an i.i.d. standard MNTS as an assumptive distribution that $\eta_t$ follows. We also hypothesize that $\eta_t$ follows an i.i.d. standard normal and student t as competing models. Hereafter, we denote the ARMA(1,1)-GARCH(1,1) model with multivariate normal, student t, and NTS distributed standard residuals as AGNormal, AGT, and AGNTS, respectively. 

We prefer the MNTS to other miscellaneous non-Gaussian distributions because of its flexibility with respect to a multivariate extension. Both the estimation of the MNTS from real data and the scenario generation based on the estimated MNTS are feasible without computational difficulty even in considerably high dimensional settings. These features are critical in their application to portfolio optimization. Also, the reproductive property of the stable distribution has an affinity for portfolio modeling.

\subsection{Risk measures}

We introduce FH risk. Let a gamble be any bounded random variable $g$ with a positive expected value and a positive probability of losses: $\mathbb{E}(g)>0$, $\mathbb{P}(g<0)>0$. FH risk is the minimum reserve that an agent should initially possess to prevent himself from almost certainly going bankrupt, even after the infinite repetition of the gamble $g$. \citet{Foster:2009} demonstrate that, for a gamble $g$, irrespective of the utility function, FH risk $R(g)$ is the unique positive root of the following equation:
\begin{equation}
\label{eq:def_FH}
\mathbb{E}\left(\log\left[1+\dfrac{g}{R(g)}\right]\right)=0.
\end{equation}
The bankruptcy-proof property endows FH risk with extremely high sensitivity to negative events. By regarding investments in financial assets as a gamble, FH risk is expected to sense market downturn in a forward-looking manner.

We also utilize more popular risk measures, Value at Risk (VaR) and Average VaR (AVaR), in order to supplement FH risk.\footnote{AVaR is also called Conditional VaR or Expected Shortfall.} Accuracy of risk forecasting is an important aspect of time series models. Statistically backtesting VaR and AVaR is feasible due to their relative simplicity, whereas no backtesting methodology has been established for FH risk. We backtest VaR by the Christoffersen's likelihood ratio (CLR) test as well as AVaR by the Berkowitz's likelihood ratio (BLR) tail test and Acerbi and Szekely (AS) test. See \cite{Christoffersen:1998}, \cite{Berkowitz:2001}, and \cite{Acerbi:2014}, respectively.

\section{Data and estimation}
\label{Data and Estimation}

Our dataset contains daily logarithmic returns of cryptocurrency exchange spot rates in U.S. Dollars per unit from 2015/08/31 to 2020/03/31, resulting in 1,674 observations for each cryptocurrency. Following \cite{Caporale:2019}, we select the following four cryptocurrencies as samples: Bitcoin (BTC), Ethereum (ETH), Litecoin (LTC), and XRP. The data source is CoinMarketCap.\footnote{\url{https://coinmarketcap.com/}} Table \ref{table:DS} reports the descriptive statistics of our dataset. All cryptocurrencies have larger kurtosis than those of the normal distribution, and show either negative or positive skewness. These observations motivate us to apply the non-Gaussian model. We estimate all models based on maximum likelihood estimation. In order to obtain the AGNTS model, we estimate the univariate AGT model for each cryptocurrency first and subsequently fit the standard MNTS distribution to the same residuals $\eta_t$. We refer the readers to \cite{Kurosaki:2019} for details regarding these techniques.

Our analysis procedure is as follows. First, we arrange a moving window with a length of 500 days. The first window, ranging from 2015/08/31 to 2017/01/12, moves ahead day by day until 2020/03/31, which amounts to 1,175 distinct windows. Subsequently, we iteratively estimate time series models from returns data within each window, and forecast a one-day-ahead return distribution. Finally, we assess the statistical performance of the resulting 1,175 models, as well as the profitability of the optimized portfolio based on a forecasted return distribution.

\section{Statistical Tests}
\label{Statistical Tests}

We assess the capability of our multivariate GARCH-type models to account for marginal return dynamics of each cryptocurrency. Specifically, we examine the statistical performance of the 1,175 iteratively-estimated models based both on in-sample and out-of-sample tests.
 
\subsection{In-Sample Test}
 
As an in-sample test, we investigate the fitting performance of standard residuals $\eta_t$ of the univariate ARMA(1,1)-GARCH(1,1) model for the assumptive distributions (normal, student t, and NTS). To do so, we exploit both the Kolmogorov-Smirnov (KS) and the Anderson-Darling (AD) tests. While both tests are designed to assess the goodness-of-fit of the proposed distributions, the AD test puts more emphasis on fitting at the tail. Under the reasonable postulation that our sample is sufficient in number, we can compute p-values for both tests.

Tables \ref{table:KS} and \ref{table:AD} report the number of rejections of KS and AD tests out of 1,175 iterated estimations for AGNormal, AGT, and AGNTS residuals, respectively, by significance level. We see that AGNormal is almost always rejected by both tests and thus significantly underperforms AGT and AGNTS. In the KS test, AGNTS has a smaller number of rejections than AGT in three out of four cryptocurrencies at the 10\% level. More clearly, in the AD test, AGNTS has a smaller number of rejections than AGT in four (three) out of four cryptocurrencies at the 5\% (10\%) level due to the excellent ability of AGNTS to track tail behavior. Overall, AGNTS is the most preferable model.

\subsection{Out-of-Sample Test}

As an out-of-sample test, we backtest risk measures, namely, VaR and AVaR. Each of the iteratively-estimated models forecasts one-day-ahead VaR and AVaR, constituting a time series of VaR and AVaR forecasts with a length of 1,175 days from 2017/01/12 to 2020/03/31.\footnote{Following \citet{Kim:2010}, the VaR estimation with AGNTS relies on the discrete Fourier Transform.} In line with the Basel accord, we adopt the 99\% confidence level for VaR and AVaR. Backtesting is achieved by comparing VaR and AVaR with actual returns every day. Out-of-sample testing is more important than in-sample testing since our research interest lies in portfolio risk forecasting and optimization. In order to clarify our results, we divide the sample period into three subperiods and conduct out-of-sample tests by subperiod. Specifically, let Periods 1, 2, and 3 cover from 2017/01/12 to 2018/03/31, from 2018/04/01 to 2019/03/31, and from 2019/04/01 to 2020/03/31, respectively. Notice that each subperiod includes some form of turmoil. The cryptocurrency boom and crash around the end of 2017 for Period 1, the crash following the release of "hardfork" of Bitcoin cash at the end of 2018 for Period 2, and the Covid-19 crisis in the beginning of 2020 for Period 3. 

Table \ref{table:out-of-sample-tests} summarizes the p-values of out-of-sample tests. The CLR test with conditional coverage is for VaR forecasts, and the BLR tail and the AS tests are for AVaR forecasts.\footnote{The p-values of the AS test are computed from $10^4$ sample statistics generated by time series models and for a left tail.} First of all, AGNormal has lower p-values than AGT and AGNTS, especially in the BLR and AS tests. AGNTS passes the CLR tests during any period and with any cryptocurrency, including at the 10\% level, whereas AGT fails in Periods 1 and 2 at the same level. Also, AGNTS always passes the BLR tests except for in Period 2 and in BTC. By contrast, AGT fails in Periods 1 and 2 at the 5\% level. Finally, AGNTS fails the AS tests for at most one cryptocurrency in each subperiod at the 10\% level. However, AGT fails for two cryptocurrencies in Periods 1 and 3 at the same level. Therefore, we conclude that AGNTS shows the best performance in out-of-sample tests more clearly than in in-sample tests.

\section{Portfolio optimization}
\label{Portfolio Optimization}

We practice portfolio optimization with cryptocurrencies consisting of BTC, ETH, LTC, and XPR, in line with \citet{Kurosaki:2019}. The portfolio risk and reward is forecasted through multivariate time series models. The optimization is carried out by minimizing the objective risk measure under the tradeoff with expected returns and transaction costs following the procedure detailed in \ref{Portfolio Optimization Procedure}. Since, as is shown in Section \ref{Statistical Tests}, AGNTS provides a more accurate expression than AGNormal and AGT for leptokurtic and skewed behaviors of cryptocurrency returns, it is also expected to show better performance in portfolio optimization. We exploit FH risk as the objective risk measure to be minimized, as well as standard deviation (SD) and AVaR.

Table \ref{table:performance} exhibits optimization results under the absence of transaction costs ($\lambda=0$), through a combination of time series models and objective risk measures. When the portfolio is optimized with respect to SD, AVaR, and FH under the tradeoff against expected returns, we refer to the corresponding portfolio as mean-SD, mean-AVaR, and mean-FH portfolio, respectively. Column 2 reports the cumulative returns that each optimized portfolio accrues during the period from 2017/01/12 to 2020/03/31.\footnote{Since AGT and AGNTS share the same residuals, both models produce the same mean-SD portfolio. Also, note that the first revenue recognition takes place on the day after the first optimization, 2017/01/13.} We see that the mean-FH portfolio with AGNTS forecasts yields the largest profit, followed by the mean-AVaR portfolio with AGNTS forecasts. Columns 3 through 5 show the SD, AVaR, and FH risk of the optimized portfolio itself, which are computed from their historical returns. Columns 6 through 8 indicate the cumulative returns adjusted by each risk measure. Note that the return-to-SD ratio in Column 6 is the well-known Sharpe ratio. We find that the mean-FH portfolio with AGNTS forecasts shows the highest return-to-risk ratios of any risk measures. Therefore, we conclude that this combination not only achieves the largest cumulative returns but also generates the most ideal balance between risk and reward.

\begin{figure}[htbp]
	\begin{center}
		\begin{subfigmatrix}{2}
			\subfigure[$\lambda=0$]{\includegraphics[width=0.45\linewidth]{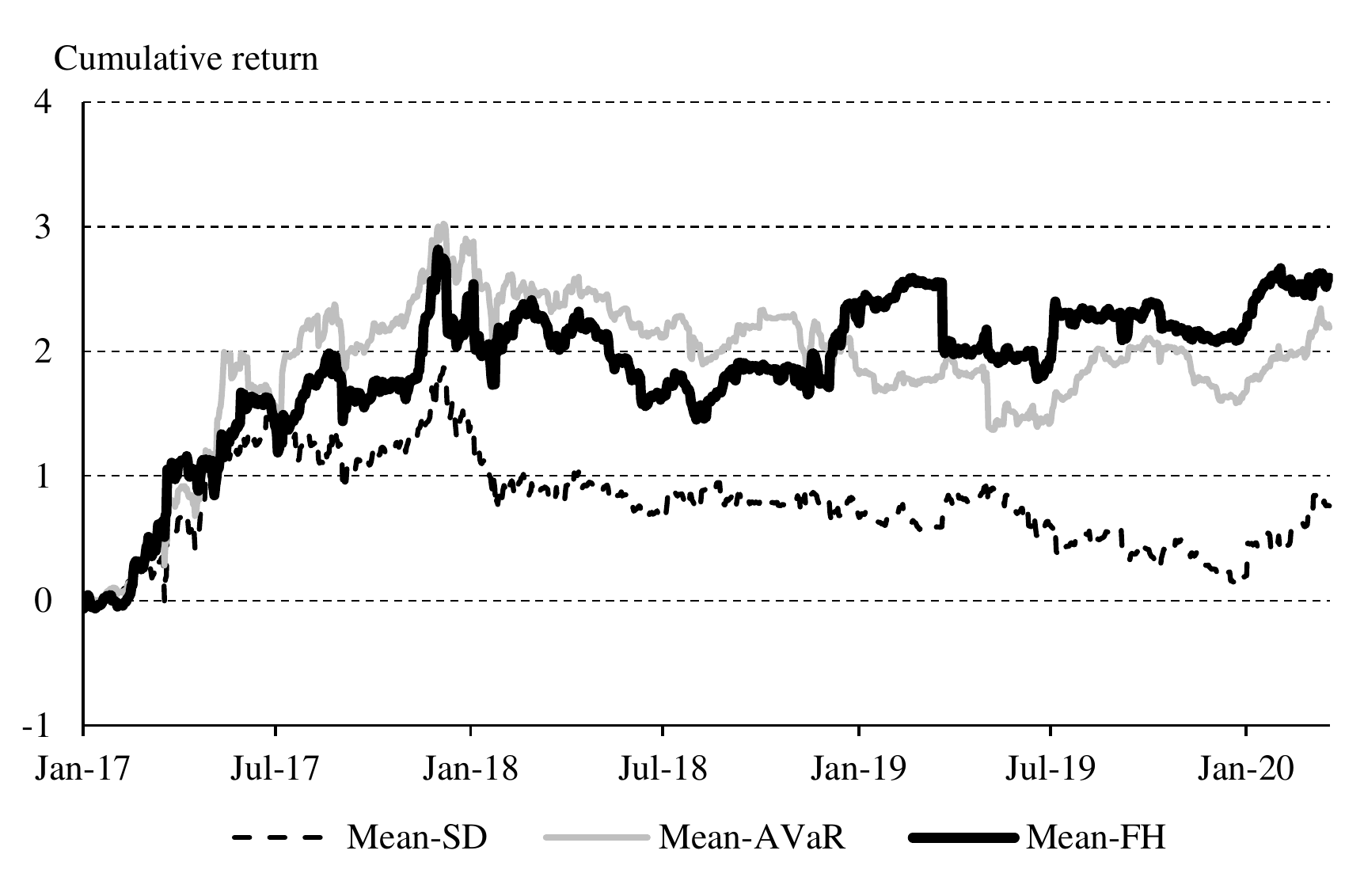}\label{fig1a}}
			\subfigure[$C=0.01$]{\includegraphics[width=0.45\linewidth]{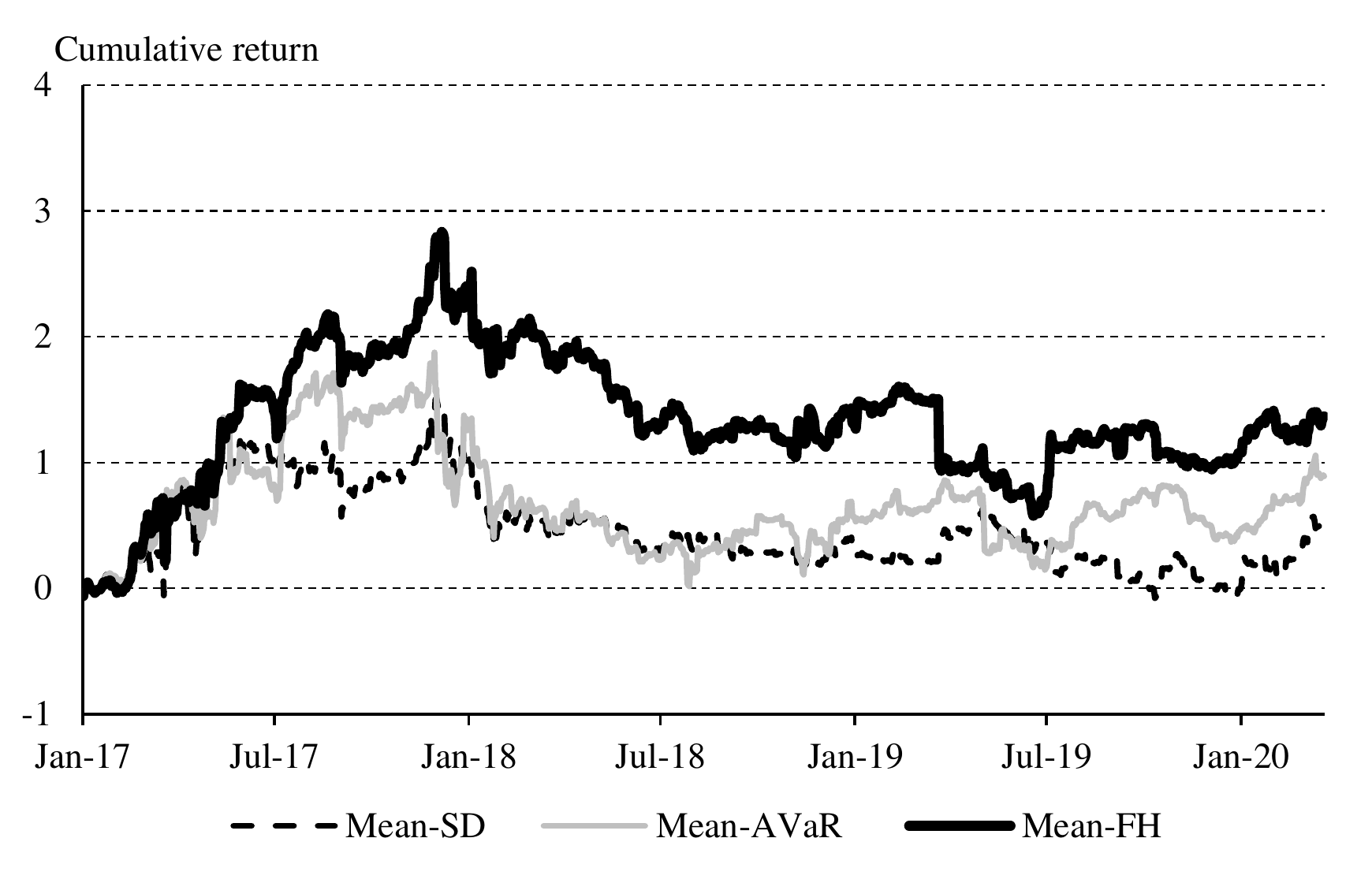}\label{fig1b}}
			\subfigure[$C=0.1$]{\includegraphics[width=0.45\linewidth]{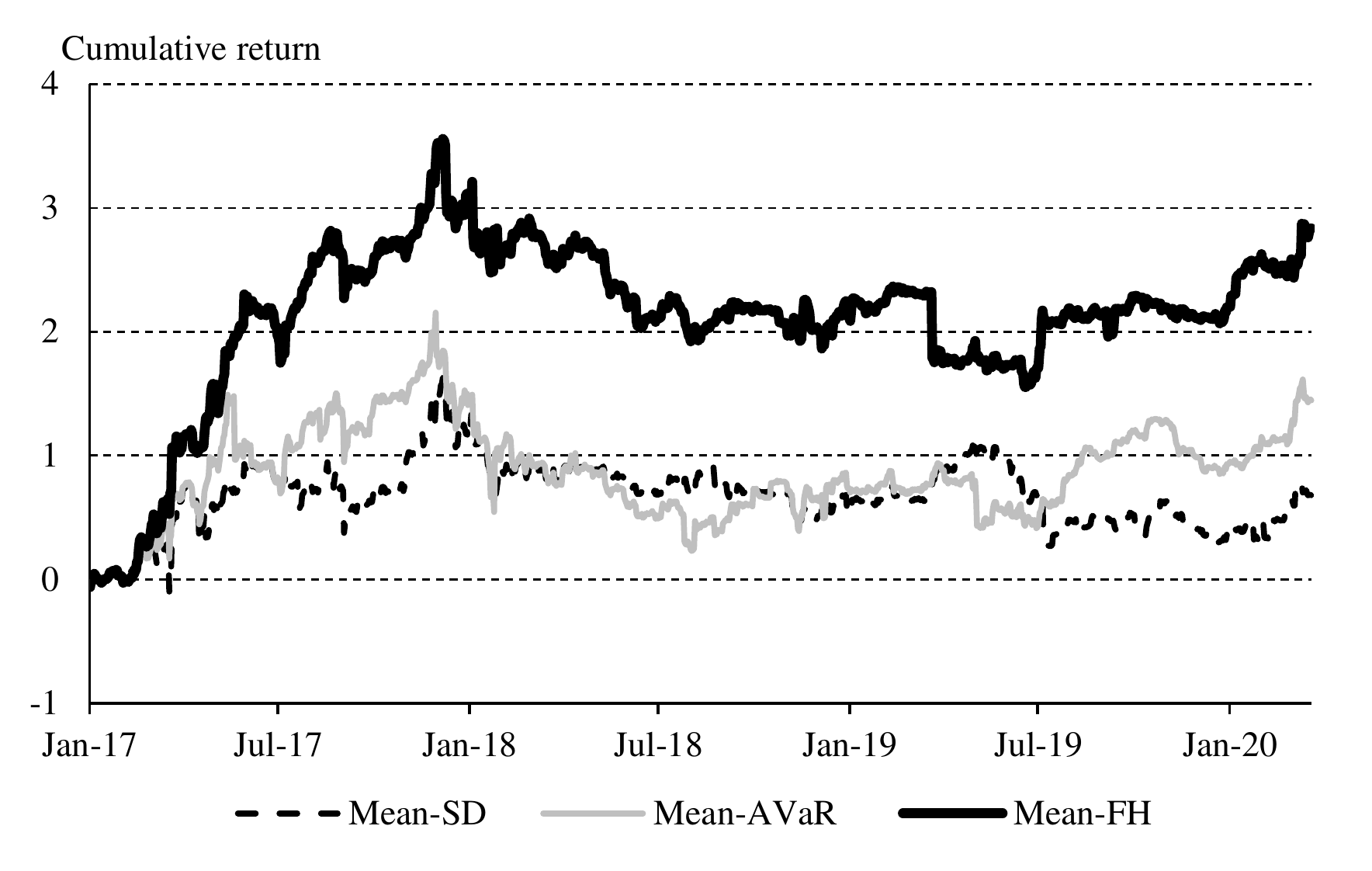}\label{fig1c}}
			\subfigure[$C=1$]{\includegraphics[width=0.45\linewidth]{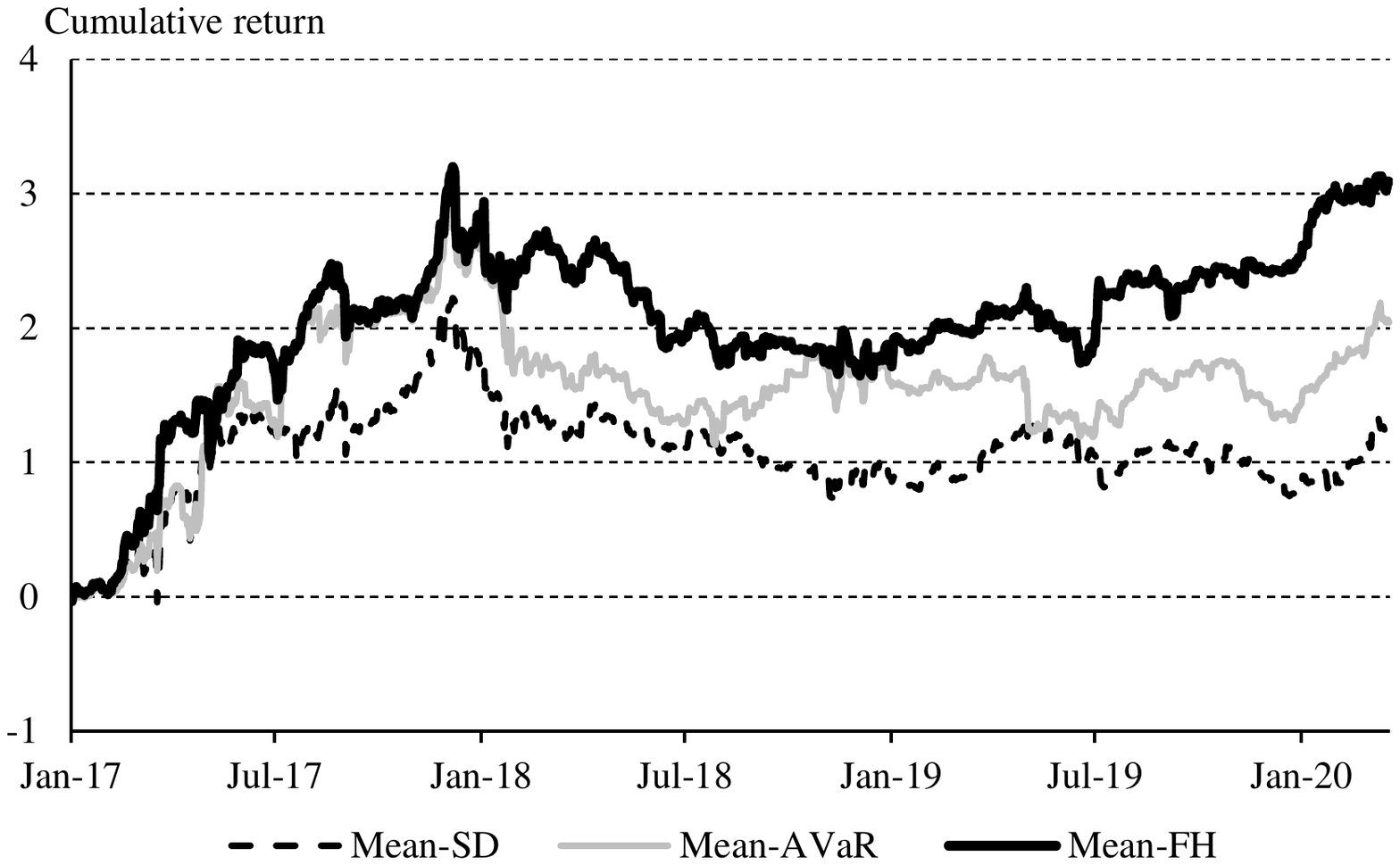}\label{fig1d}}
		\end{subfigmatrix}
	\end{center}
	\caption{Time Evolution of Optimized Portfolio's Cumulative Returns.
	}
	\label{fig:Time_evolution_return}
\end{figure}

As a robustness check, we conduct the optimization in the presence of transaction costs, where the corresponding cost ($\lambda\cdot |w_t-w_{t-1}|$) is deducted from daily returns. Figure \ref{fig:Time_evolution_return} shows how the cumulative returns of the optimized portfolio evolve temporarily over the investment period based on different investor cost aversion ($C=0.01, 0.1, 1$), as well as in cases without transaction costs ($\lambda=0$). We focus on the forecasts of future returns created by AGNTS. We observe that the mean-FH portfolio accrues the largest profit irrespective of cost aversion parameters. Therefore, the superiority of the combination of AGNTS and FH risk still holds for the case where the transaction cost is present.\footnote{The presented optimization allows for both long and short positions. As a further robustness check, we conduct the optimization under the restriction that a short position is prohibited. This restriction is realistic for conservative investors. The results are consistent with unrestricted cases; The mean-FH portfolio and/or the AGNTS forecasts generally create the largest profit.}

\section{Concluding remarks}
\label{Concluding Remarks}

This paper studies the portfolio optimization of four major cryptocurrencies. A cryptocurrency as an asset class is characterized by higher volatility and more nonlinear return dynamics compared to traditional assets. Statistical analysis demonstrates that the introduction of the MNTS distribution enhances the explanatory power of the GARCH-type model for cryptocurrency return dynamics substantially, especially in terms of risk forecasting. FH risk warns of market crashes, for example, in scenarios such as those caused by the Covid-19 pandemic, in an immeasurably sensitive manner. The combination of the MNTS distributed GARCH model and FH risk leads to desirable portfolio optimization with respect to cumulative returns as well as risk-return balance. We first document the effectiveness of those sophisticated techniques in the context of cryptocurrency.




\bibliographystyle{elsarticle-harv}
\bibliography{FH_cryptocurrency_arxiv_bib}

\clearpage

\begin{table}[htbp]
	\begin{center}
		\caption{Descriptive statistics of our dataset. The sample period is from 2015/08/31 to 2020/03/31.}
		\label{table:DS}
		\vspace{10pt}
		\scalebox{0.8}
		{
			\begin{tabular}{cccccccc}\hline
				\raisebox{4pt}{Cryptocurrency} & \shortstack{Number of \\ observation} & \raisebox{4pt}{Mean} & \raisebox{4pt}{Max} & \raisebox{4pt}{Min} &
				\shortstack{Standard \\ deviation} & \raisebox{4pt}{Kurtosis} & \raisebox{4pt}{Skewness} \\\hline
				BTC & 1674 & 0.0020  & 0.2251  & $-0.4647$  & 0.0405  & 13.9441  & $-0.9392$  \\
				ETH & 1674 & 0.0027  & 0.3028  & $-0.5507$  & 0.0624  & 7.0200  & $-0.1820$  \\
				LTC & 1674 & 0.0016  & 0.5103  & $-0.4490$  & 0.0565  & 12.8712  & 0.8100  \\
				XRP & 1674 & 0.0019  & 1.0274  & $-0.6163$  & 0.0689  & 43.1126  & 2.8848  \\\hline
			\end{tabular}
		}
	\end{center}
\end{table}


\begin{table}[htbp]
	\begin{center}
		\caption{Number of rejections of Kolmogorov-Smirnov tests out of 1,175 iterated estimations for GARCH residuals}
		\label{table:KS}
		\vspace{10pt}
		\scalebox{0.8}
		{
			\begin{tabular}{cccccccccc}
				\shortstack{Significance \\ level} & \multicolumn{3}{c}{\raisebox{10pt}{1\%}} & \multicolumn{3}{c}{\raisebox{10pt}{5\%}} & \multicolumn{3}{c}{\raisebox{10pt}{10\%}} \\\hline
				Model & AGNormal & AGT & AGNTS & AGNormal & AGT & AGNTS & AGNormal & AGT & AGNTS \\\hline
				BTC & 1009 & 92 & 107 & 1079 & 186 & 185 & 1163 & 303 & 259 \\
				ETH & 1110 & 136 & 134 & 1173 & 337 & 383 & 1175 & 542 & 523 \\
				LTC & 889 & 416 & 488 & 1167 & 532 & 577 & 1175 & 648 & 693 \\
				XRP & 1174 & 683 & 650 & 1175 & 783 & 764 & 1175 & 812 & 791 \\\hline
			\end{tabular}
		}
	\end{center}
\end{table}

\begin{table}[htbp]
	\begin{center}
		\caption{Number of rejections of Anderson-Darling tests out of 1,175 iterated estimations for GARCH residuals}
		\label{table:AD}
		\vspace{10pt}
		\scalebox{0.8}
		{
			\begin{tabular}{cccccccccc}
				\shortstack{Significance \\ level} & \multicolumn{3}{c}{\raisebox{10pt}{1\%}} & \multicolumn{3}{c}{\raisebox{10pt}{5\%}} & \multicolumn{3}{c}{\raisebox{10pt}{10\%}} \\\hline
				Model & AGNormal & AGT & AGNTS & AGNormal & AGT & AGNTS & AGNormal & AGT & AGNTS \\\hline
				BTC & 1022 & 137 & 81 & 1175 & 537 & 155 & 1175 & 629 & 236 \\
				ETH & 1142 & 106 & 103 & 1175 & 329 & 287 & 1175 & 503 & 523 \\
				LTC & 1144 & 377 & 462 & 1175 & 544 & 543 & 1175 & 665 & 648 \\
				XRP & 1175 & 557 & 729 & 1175 & 812 & 801 & 1175 & 835 & 821 \\\hline
			\end{tabular}
		}
	\end{center}
\end{table}

\begin{table}[htbp]
	\begin{center}
		\caption{p-values of Three Out-of-Sample Tests}
		\label{table:out-of-sample-tests}
		\scalebox{0.8}
		{
			\begin{tabular}{cccccccccc} 				
				Test & \multicolumn{3}{c}{CLR} & \multicolumn{3}{c}{BLR} & \multicolumn{3}{c}{AS} \\\hline
				Model & AGNormal & AGT & AGNTS & AGNormal & AGT & AGNTS & AGNormal & AGT & AGNTS \\\hline
				\multicolumn{9}{c}{Period 1 (2017/01/12 to 2018/03/31)} \\\hline
				BTC & 0.0020  & 0.0174  & 0.7307  & 0.0003  & 0.0338  & 0.7824  & 0.0000  & 0.0186  & 0.6921  \\
				ETH & 0.0450  & 0.6691  & 0.6691  & 0.0000  & 0.6304  & 0.8778  & 0.0000  & 0.5016  & 0.3827  \\
				LTC & 0.4489  & 0.4489  & 0.4489  & 0.0009  & 0.7565  & 0.7650  & 0.5963  & 0.8748  & 0.8455  \\
				XRP & 0.0450  & 0.2217  & 0.4153  & 0.0000  & 0.5008  & 0.5698  & 0.0000  & 0.0835  & 0.0510  \\\hline
				\shortstack{\\ Number of p-values \\ less than 5\%} & 3 & 1 & 0 & 4 & 1 & 0 & 3 & 1 & 0 \\
				\shortstack{\\ Number of p-values \\ less than 10\%} & 3 & 1 & 0 & 4 & 1 & 0 & 3 & 2 & 1 \\\hline
				\multicolumn{9}{c}{Period 2 (2018/04/01 to 2019/03/31)} \\\hline
				BTC & 0.0388  & 0.0990  & 0.2236  & 0.0000  & 0.1242  & 0.0546  & 0.0000  & 0.0014  & 0.0001  \\
				ETH & 0.0137  & 0.4419  & 0.7271  & 0.0000  & 0.3352  & 0.8721  & 0.0000  & 0.1670  & 0.2896  \\
				LTC & 0.4419  & 0.2236  & 0.4419  & 0.0872  & 0.0064  & 0.1640  & 0.0319  & 0.2083  & 0.2491  \\
				XRP & 0.7271  & 0.3366  & 0.3366  & 0.5057  & 0.3174  & 0.5176  & 0.2599  & 0.9671  & 0.9657  \\\hline
				\shortstack{\\ Number of p-values \\ less than 5\%} & 2 & 0 & 0 & 2 & 1 & 0 & 3 & 1 & 1 \\
				\shortstack{\\ Number of p-values \\ less than 10\%} & 2 & 1 & 0 & 3 & 1 & 1 & 3 & 1 & 1 \\\hline
				\multicolumn{9}{c}{Period 3 (2019/04/01 to 2020/03/31)} \\\hline
				BTC & 0.0394  & 0.4449  & 0.7295  & 0.0000  & 0.4436  & 0.2497  & 0.0000  & 0.0177  & 0.0420  \\
				ETH & 0.1002  & 0.7295  & 0.6733  & 0.0000  & 0.9374  & 0.8386  & 0.0000  & 0.3157  & 0.6138  \\
				LTC & 0.1002  & 0.4449  & 0.7295  & 0.0000  & 0.9158  & 0.9124  & 0.0000  & 0.1602  & 0.2592  \\
				XRP & 0.2258  & 0.2258  & 0.4449  & 0.0000  & 0.5051  & 0.7859  & 0.0000  & 0.0731  & 0.1274  \\\hline
				\shortstack{\\ Number of p-values \\ less than 5\%} & 1 & 0 & 0 & 4 & 0 & 0 & 4 & 1 & 1 \\
				\shortstack{\\ Number of p-values \\ less than 10\%} & 1 & 0 & 0 & 4 & 0 & 0 & 4 & 2 & 1 \\\hline
			\end{tabular}
		}
	\end{center}
\end{table}

\begin{table}[htpb]
	\begin{center}
		\caption{Performance of Each Optimized Portfolio ($\lambda=0$)}
		\label{table:performance}
		\vspace{10pt}
		\scalebox{0.8}
		{
			\begin{tabular}{cccccccc} 	
				Portfolio & \shortstack{Cumulative \\ Return (a)} & SD (b) & AVaR (c) & FH risk (d) & a/b & a/c & a/d \\\hline
				\multicolumn{8}{c}{AGNormal}  \\\hline
				Mean-SD & 0.8662 & 0.0415 & 0.2293 & 1.2554 & 20.8968 & 3.7773 & 0.6900 \\
				Mean-AVaR & $-0.3812$ & 0.0464 & 0.2766 & 0.4555 & $-8.2220$ & $-1.3785$ & $-0.8371$ \\
				Mean-FH & 0.0677 & 0.0509 & 0.2761 & 22.5218 & 1.3288 & 0.2450 & 0.0030 \\\hline
				\multicolumn{8}{c}{AGT}  \\\hline
				Mean-SD & 0.7612 & 0.0427 & 0.1981 & 1.4200 & 17.8362 & 3.8424 & 0.5361 \\
				Mean-AVaR & 1.4448 & 0.0465 & 0.2392 & 0.9514 & 31.0754 & 6.0402 & 1.5186 \\
				Mean-FH & 1.7895 & 0.0673 & 0.3590 & 1.6465 & 26.6043 & 4.9845 & 1.0869 \\\hline
				\multicolumn{8}{c}{AGNTS}  \\\hline
				Mean-SD & 0.7612 & 0.0427 & 0.1981 & 1.4200 & 17.8362 & 3.8424 & 0.5361 \\
				Mean-AVaR & 2.1916 & 0.0488 & 0.2347 & 0.6627 & 44.9341 & 9.3391 & 3.3072 \\
				Mean-FH & \textbf{2.5889} & 0.0527 & 0.2518 & 0.7347 & \textbf{49.0979} & \textbf{10.2808} & \textbf{3.5235} \\\hline
			\end{tabular}
		}
	\end{center}
\end{table}

\newpage


\clearpage

\appendix

\section{Supplement to Methodology}
\label{Supplement to Methodology}

\subsection{The MNTS distribution}

A real parameter set $(\alpha,\theta,\beta,\gamma,\mu,\Sigma)$ characterizes the $n$-dimensional MNTS distribution, where $\alpha\in(0,2)$, $\theta>0$, $\beta, \mu\in \mathbb{R}^n$, $\gamma \in \mathbb{R}^n_+$, and $\Sigma$ is a $n$-by-$n$ correlation matrix \citep{Kim:2012}. The common parameters $(\alpha,\theta)$ control fat-tailness. Each component of $\beta$, $\gamma$, and $\mu$ corresponds to each marginal; $\beta$ controls skewness, $\gamma_i$ scales the distribution, and $\mu$ is a mean value. The correlation matrix $\Sigma$ is responsible for asymmetric interdependence among marginals. These unique parameters enable the MNTS to have explanatory powers for the stylized nature of financial asset returns, such as fat-tailness and skewness, which a normal distribution fails to capture. See \cite{Kim:2012} for more detail.

\subsection{VaR and AVaR}

The VaR at the $1-\epsilon$ confidence level is the $\epsilon$-quantile of an asset loss. AVaR is the expected loss on the condition that it is at levels in excess of VaR. As well as FH risk, AVaR is a convex function with a unique minimum. Therefore, the portfolio can be effectively optimized with respect to AVaR and FH risk. This is not the case with VaR because it generally has multiple local minima. It is also noteworthy that VaR and AVaR should be defined with time horizon and confidence level, whereas FH risk is an absolute risk measure irrespective of such predetermined parameters.

The Christoffersen's likelihood ratio (CLR) test is a well-known method for backtesting VaR as an interval forecast \citep{Christoffersen:1998}. More specifically, we basktest VaR through a CLR test with a conditional coverage property, which is the joint test of unconditional coverage and independence, because it can take the tendency for consecutive VaR breaches into consideration. Furthermore, we backtest AVaR both directly and indirectly. For the indirect approach, we utilize a Berkowitz's likelihood ratio (BLR) tail test \citep{Berkowitz:2001}. The BLR tail test can backtest the accuracy of tail behavior forecasts of a given distribution. Accurate tail behavior forecasts coincide with accurate AVaR forecasts. For our direct approach, we exploit the Acerbi and Szekely (AS) test \citep{Acerbi:2014}. While they propose several statistics so as to backtest the estimated AVaR at the $1-\epsilon$ confidence level, we choose the following statistic:
\begin{equation}
\label{eq:def_AS_statistics}
Z = \sum_{t=1}^T \frac{R_t I_t}{T\epsilon \textrm{AVaR}_{\epsilon, t}} +1,
\end{equation}
where $R_t$ is a historical asset return, $I_t$ is the indicating function for VaR breaches, i.e., $I_t=1_{-R_t>\textrm{VaR}_{\epsilon, t}}$, $T$ is the forecast length, and $t$ is the time period. If the AVaR forecast is accurate (the null hypothesis), $Z$ must be zero; otherwise, $Z$ is negative. The p-values of the AS test can be derived by simulating $R_t$ through a time series model and sampling Z.

\section{Portfolio Optimization Procedure}
\label{Portfolio Optimization Procedure}

The presented problem is to search for the optimized weight of each cryptocurrency $i$ with expected return $\mu_{i,t}$ at time $t$, denoted by $w_{i,t}$ $(1\le i\le4)$, conditional on the information available up to time $t$, under the tradeoff between risk and reward. We identify this problem as minimizing the risk-to-reward ratio (disutility), where the risk and the reward are quantified by FH risk and expected returns, respectively. We supplementarily utilize one-day-ahead AVaR at the 99\% confidence level and standard deviation (SD) as risk measures for reference. We exclude VaR because of its non-convexity. Note that the optimization based on SD is equivalent to the classical Markowitz framework.

We impose some constraints on weights $w_{i,t}$. First, we allow for both long and short positions up to unit, since short positions are typical in cryptocurrency trading especially after the issuance of Bitcoin futures in December, 2017. Second, we maintain positive expected returns from the portfolio. The second constraint stems from the principle of speculation as well as the necessary condition for FH risk to be defined. In addition, we consider the cost of reallocating portfolio weights, which partly deprives the portfolio of cumulative returns. We also take investor cost aversion into account.

Consequently, our optimization problem is described using the following equation:
\begin{equation}
\label{eq:optimization}
\min_{\substack{-1\le w_{i,t}\le 1 \\ 0\le w_t^{\top}\mu_t}} \dfrac{\rho(w_t)}{w_t^{\top}\mu_t} + C \left[ \dfrac{\lambda\cdot (w_t-w_{t-1})}{w_t^{\top}\mu_t} \right]^2,
\end{equation}
where $w_t=(w_{1,t},\hdots,w_{4,t})^{\top}$, $\mu_t=(\mu_{1,t},\hdots,\mu_{4,t})^{\top}$, and $\rho(\cdot)$ is a risk measure of the portfolio. The second quadratic term of the change in weights corresponds to the transaction cost \citep[e.g.,][]{Fabozzi:2006}. $\lambda$ and $C$ are the parameters for transaction cost charged per unit weight change and investor cost aversion relative to risk-to-reward ratio, respectively. We set $\lambda$ as $10^{-7}$ when transaction costs are present.\footnote{Since $\mu_{i,t}$ is typically on the order of $10^{-3}$ in our dataset, $\lambda=10^{-7}$ suggests that transaction costs are roughly on the order of 1 bps to the portfolio expected returns.} We assume several distinct values for $C$ since it depends on investors. Notice that our objective function (\ref{eq:optimization}) is set as homogeneous with respect to the size of $w_t$, as long as $\rho(\cdot)$ satisfies the homogeneity.

We therefore have the 1,175 iteratively-estimated time series models from 2017/01/12 to 2020/03/31. Every day during this investment period, we generate one-day-ahead $10^4$ scenarios for each cryptocurrency return by utilizing the estimated AGNormal, AGT, and AGNTS models. Under these generated scenarios, we forecast risk and reward, thereby finding the optimal portfolio $w_t$ to solve equation (\ref{eq:optimization}). The portfolio is optimized into $w_t$ at the end of day $t$ and create profit or loss from the return at the end of day $t+1$.

\end{document}